\RequirePackage{fix-cm}
\documentclass[smallextended]{svjour3}       
\smartqed  
\usepackage{graphicx}
\usepackage{amsmath}
\usepackage{array}
\usepackage{amssymb}
\usepackage{bm}
\usepackage{paralist}
\usepackage{graphics} 
\usepackage{epsfig} 

 \journalname{Netnomics}
\begin{document}

\title{Can social microblogging be used to forecast intraday exchange rates?
}


\author{Panagiotis Papaioannou \and
        Lucia Russo  \and
        George Papaioannou \and
        Constantinos I. Siettos 
}


\institute{P. Papaioannou \at
              School of Applied Mathematics and Physical Sciences\\
              National Technical University of Athens, Greece\\
           \and
            L. Russo \at
            National Research Council, Naples, Italy\\
                \and
                 G. Papaioannou \at
              Center for Research and Applications of Nonlinear Systems\\
              CRANS, University of Patras, and ADMIE,Greece\\
              \and
            C. I. Siettos \at
            School of Applied Mathematics and Physical Sciences \\
              National Technical University of Athens, Greece\\
              \email{ksiet@mail.ntua.gr}
              }

\date{ This is a prior version of the paper published at NETNOMICS. The final publication is available at http://www.springer.com/economics/economic+theory/journal/11066}

\maketitle

\begin{abstract}
The Efficient Market Hypothesis (EMH) is widely accepted to hold
true under certain assumptions. One of its implications is that the
prediction of stock prices at least in the short run cannot
outperform the random walk model. Yet, recently many studies
stressing the psychological and social dimension of financial
behavior have challenged the validity of the EMH. Towards this aim,
over the last few years, internet-based communication platforms and
search engines have been used to extract early indicators of social
and economic trends. Here, we used Twitter's social networking
platform to model and forecast the EUR/USD exchange rate in a
high-frequency intradaily trading scale. Using time series and
trading simulations analysis, we provide some evidence that the
information provided in social microblogging platforms such as
Twitter can in certain cases enhance the forecasting efficiency
regarding the very short (intradaily) forex.

\end{abstract}

\section{Introduction}
\label{intro} The exchange rate forecasting is one of the most
significant, yet tough research pursuits of contemporary financial
management. Volatility risk is directly connected not only to
company but also to national and international-level macroeconomic
relations and strategic measures. Hence, it is not a surprise that
markets and organizations such as the Federal Reserve have spent an
inordinate amount of both time and money in trying to develop models
able to accurately predict the future. Over the years, studies have
proceeded mainly on two fronts. On one hand, there are the
fundamental models trying to project the exchange rates based on
rational expectations hypotheses involving major macroeconomical
figures such as national incomes, expected inflation differentials,
supplies and demands of the exchanged currencies. This category
includes models based on the purchasing power parity (Keneth, 1996),
covered and uncovered interest rate parity (Chaboud and Wright,
2005; Chinn et al., 2004) and monetary models (Frankel, 1982;
MacDonald and Taylor, 1994; Groen, 2000). However, as Richard Meese
and Kenneth Rogoff showed back in 1983 (Meese et al., 1983), such
structural models cannot outperform the forecasting capability of a
naive random-walk at least in the short run. On the other hand,
there are the so-called unstructured models which use time-series
statistics to predict currency movements. This category includes
regression models (Huang et al., 2005; Preminger and Franck, 2007),
Markov models (Mamon and Elliott, 2007; Park, et al., 2009;
Shmilovici et al., 2009; Nikolsko-Rzhevskyy and Prodan, 2011),
support vector regression (Burges, 1998; Van Gestel et al., 2001;
Tay and Cao, 2002; Kim 2003; Huanga et al., 2010), artificial neural
networks and genetic algorithms (Kuan and Liu 1995; Yao and Tan,
2000; Liao and Tsao, 2006). Recently, various agent-based models
based on behavioral finance concepts (Shleifer, 2000) have been
proposed that relax the standard hypothesis of homogeneous perfectly
informed agents with expectations consistent with the theoretical
ones (Steiglitz and Shapiro, 1998; Carpenter, 2002; Iori, 2002;
Marsilia and Raffaelli, 2006; Corona et al., 2008). Indeed, news
diffusion and social mimesis through social networking have been,
especially over the last few decades, primary factors in shaping not
only markets but also economical and political changes around the
globe (Garcia, 1997; Hon et al., 2007; Johansen, 2004). Under this
perspective, identifying and understanding social collective
behavior as this emerges due to individuals' interactions has become
a key element in today's economy (Camerer, 1999; Daniel et al.,
2002; Ross, 2008; Casti, 2010; Knauff et al., 2010). However, also
these models, due to the inherent extraordinary complexity of the
problem, they are built on incomplete knowledge and for that reason
they are flashing a ``note of caution" on their robustness and
efficiency.  As stated by the former Chairman of the Federal Reserve
of the United States Alan Greenspan in 2002 ``There may be more
forecasting of exchange rates, with less success, than almost any
other economic variable" (Greenspan, 2002). The efficient market
hypothesis (Fama, 1970; Milgrom and Stokey, 1982; Malkiel, 2003,
2005) has been proved by experience to hold true, at least regarding
predictions in the short run, in its two common forms: (a) the weak,
stating that future prices cannot be predicted by using any
technical analysis based on prices from the past and (b) the
semi-strong, stating that future prices cannot be predicted based on
publicly available new information such as the macroeconomic
surprises. But what about the strong form of the EMH reflecting all
kinds of information? It has been shown, that if the ``beliefs" of
the traders are concordant and the agents behave rationally, both
private and public information are valueless to speculation (Milgrom
and Stokey, 1982). However, there are studies claiming that the
celebrated Milgrom and Stokey no-trade theorem does not apply when
agents react diversely on public available information. In general,
agents exhibit heterogeneity in their behavior, they often respond
irrationally and/or diversely in the announcement of public
announcements based on their earning expectations and they are
diversely informed. Among others, the above facts have raised an
intense debate over the validity of the EMH. Regarding forex it has
been demonstrated by many studies that ``beliefs" as these are
shaped by people's private information play a major role (Bacchetta
and van Wincoop, 2006; Gyntelberg et al., 2009). But how one can
retrieve such "private" information? Nolte and Pohlmeier (2006)
analyzed the predictive capability of finance experts based on the
Centre of European Economic Research's Financial Markets Survey.
They concluded that there is no any evidence that could support the
assumption that such a survey could provide valuable information for
improving forecasting. Today, the newborn microblogging socializing
services - that have revolutionarized the way private and publicly
available information diffuses- appear as promising media to data
mining agents' personal information and ``beliefs" as these are
reflected by their (trading) behavior (Schumaker and Chen, 2009;
Asur and Huberman, 2010). For example, such services have been
exploited with the aid of search queries as tools to stock-market
prediction (Bollen et al., 2011) and movie box-office revenue (Asur
and Huberman, 2010); the modeling and prediction of other complex
phenomena such as the early detection of epidemics (Ginsberg et al.,
2009) and earthquake (Earle et al., 2010) has also been attempted.
For financial or macro-economic time series prediction, three
general categories of online sources have been exploited (see Mao
and Bollen (2011) for a review), namely News Media, Web Search (such
as Google Insight) and Social Microblogging (such as Facebook and
Twitter). These studies try to form sentiment indicators based on
keyword finding and proper interpretation. Here we follow another
path in exploiting Twitter's online data sets: we make use of
traders ``beliefs" as reflected through their published limit orders
in the Twitter. Several on-line algorithmic brokerage firms (e.g.
Zulutrade.com) publish the incoming limit orders of their retail
clients, (without displaying their identity) for other participants
to view, bid and post their own orders. It is therefore tempting to
exploit such information to enhance the forecasting potential of
exchange rates. Using various kinds of modes, namely Autoregressive
(AR), Autoregressive with exogenous input linear models (ARX) and
Artificial Neural Networks (ANN) we provide some evidence that
social microblogging services can in certain cases be used to
enhance the forecasting performance of these models in the very
short (intradaily) run.

\section{Method's summary}
\label{sec:1} Our proposed approach aims at providing evidence that
social web media such as the Twitter's microblogging platform can be
used to enhance forecasting of the exchange rate in the short run.
For our illustrations, we used a dataset of 20,250 public-available
messages posted on the Twitter's platform (with no re-tweets in
them) recorded from 25/10/2010 to 05/01/2011. Twitter launched in
2006 providing social networking through the posting of
140-character text messages among its users. Today, the estimated
daily traffic is around 65 million tweets sent by more than 190
million users. Each of these tweets was provided along with its
identifier (a username), the date and time of the tweet's
submission, and the posted text content. Using a search API on
Twitter's database, the Archivist, we searched the database in order
to match the keyword ``buy EUR/USD". Doing so, we found out that
each tweet containing the sought string was including information
about the types of orders that each Twitter user-trader had made, as
well as the target-price of each of these orders. The order types
that were posted were in their majority limit orders, that each
trader had already made, possibly through his brokerage firm, and
thus reflecting his "belief" about the upcoming EUR/USD exchange
rate quote.  Using the target-price of each message, we first
transformed each obtained number into an integer, in order to form a
solid dataset. This has been done, because many target-prices were
posted in different forms, i.e. as ``1.345" or ``1,345", ``13,45",
``134.5" etc. Due to the fact that our analysis was focused on
high-frequency intraday trading, we decided to study the temporal
behavior of the tweets in an hourly basis. As many recent financial
studies have proposed, regarding the distribution of several
financial assets (J.P. Morgan Asset Management, 2009), we found that
the intradaily tweets' -based quotes distribution follows also an
alpha-stable distribution. At this point we should note that within
our sample, there were a few days (3 out of 54 trading days) lacking
a statistically significant number of observations (due to the fact
that Twitter Archivist didn't seem to collect many tweets during
these days). To overcome the problem and just for these days, we
produced a larger sample, filling the trading hours within these
days, using the alpha-stable distribution with the same statistical
parameters of the other days (such as variance, skewness etc.),
except for the statistical mean value. This was taken to be the
Gaussian weighted-with respect to the transaction volume- mean of
the few tweets recorded in each of these days. In order to predict
the actual closing based on the tweets trend, we used a time window
of the first 50 minutes within each hour. For example, for the
tweets posted from 1 to 2.00 pm, we selected the tweets posted from
1.01 pm until 1.51 pm. Figure 1 depicts the time series of the
tweets' quotes and those of the hourly actual closing exchange
EUR/USD rates as obtained from the Yahoo Finance database.

\begin{figure*}
  \includegraphics[width=0.75\textwidth]{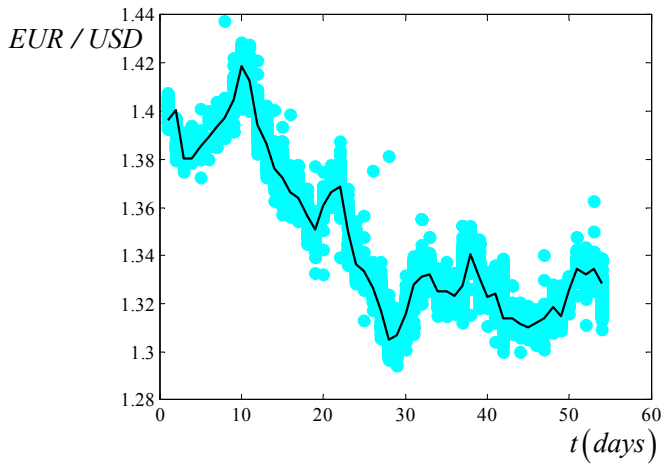}
\caption{Time series of the tweets' quotes (grey region) and those
of the hourly actual closing exchange EUR/USD rates (solid line) as
obtained from the Yahoo Finance database.}
\label{fig:1}       
\end{figure*}

By applying statistical tests (Anderson -Darling and Kolmogorov), we
found that both distributions for the total period of the 54 trading
days are hyperbolic-like distributions. More specifically, the
tweet's distribution gave a best fit to a hyperbolic distribution
with statistical mean, 1.3475 and sigma, 0.021, while for the actual
closing distribution these values were mean, 1.3488 and sigma,
0.023.

\section{The Models}
\label{sec:2}

We explored the forecasting potential of the information contained
in the tweets, and compared their prediction efficiency by
constructing (a) autoregressive (AR) (b) autoregressive exogenous
(ARX) linear models and (c)
multilayered feedforward neural networks (ANN).\\
 The general form of the AR models reads:

\begin{equation}
\hat{A}(z^{-1})y(t)=e(t)
\end{equation}

Here, $y(t)$ denotes the actual EUR/USD exchange rate at time $t$
(hourly basis); $e(t)$ is the residual at time $t$ representing the
part of the measurement that cannot be predicted from previous
measurements.\\
\begin{equation}
\hat{A}(z^{-1})=a_0+a_1z^{-1}+a_2z^{-2}+...+a_{n_a}z^{-n_a}
\end{equation}

$z^{-1}$ is the backward shift operator defined by

\begin{equation}\label{Multi}
z^{-k}y(t)=y(t-k)\\
\end{equation}

The ARX models can be written as:

\begin{equation}
y(t)=A(z^{-1})y(t-n_k)+B(z^{-1})u(t-n_k)+e(t)
\end{equation}

Here, $u(t)$ denotes the mean value of the quotes based on the
tweets as computed within the 50 minutes time interval before the
time $t$; $n_k$ is the pure time delay and

\begin{equation}\label{Multi}
\begin{split}
A(z^{-1})=a_1+a_2z^{-1}+a_3z^{-2}+...+a_{n_a}z^{-n_a}\\
B(z^{-1})=b_1+b_2z^{-1}+b_3z^{-2}+...+b_{n_b}z^{-n_b}
\end{split}
\end{equation}

For comparison purposes, we also used nonlinear regressors, namely
two-layer feedforward neural networks (ANNs). The ANNs were
constructed with two hidden layers with four nodes for each layer
and threshold functions given by $f(x)=2/(1+e^{-2x})-1$. Hence,
there are $n_a + n_b$ neurons in the input layer, and one neuron in
the output layer with a pure linear function (see Figure 2 for a
schematic of the ANN).

\begin{figure*}
  \includegraphics[width=1.0\textwidth]{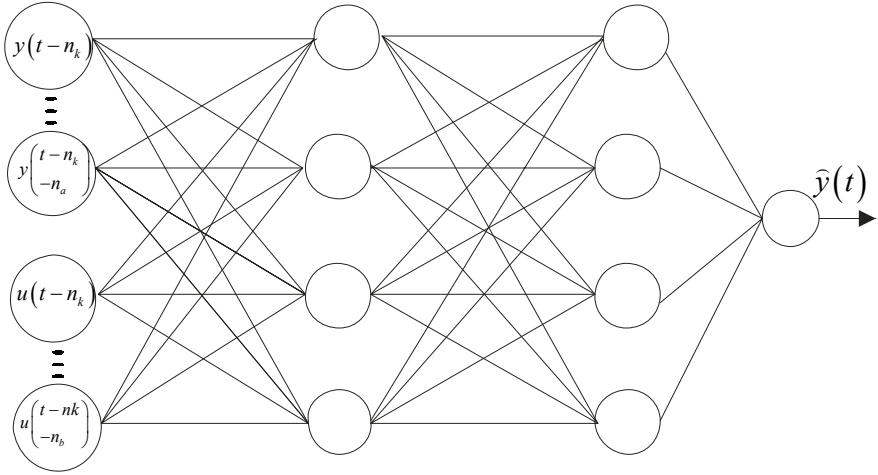}
\caption{Schematic of the ANN model}
\label{fig:2}       
\end{figure*}

The network was trained for 100 epochs with the back-propagation
algorithm based on the mean square of errors (Rumelhart et al.,
1986). Using different numbers of neurons (e.g. 3,5,6) for each
hidden layer did not change the outcomes of the analysis. For any
practical means, given the size of a training set, say, $N_T$, in
order to achieve a fair interpolation of the input space and to
avoid undesirable phenomena such as overfitting, the total number of
weights in the network, say $N_w$ should satisfy the condition $N_w
< e N$, where $e$ is the expected average approximating error (Baum
and Haussler, 1988).

The data set containing the actual closing rates and the
coarse-grained values of the tweets was split in two sets: one
containing the first $60\%$ of the data serving as a training set,
$N_T$ and the other one containing the last $40\%$ of the data
serving as a test set, say $N_V$. Different choices of the sizes of
the training and validation data sets did not change the
outcomes of the analysis. \\
The parameter estimation of both types of models was done by
least-squares fitting on the set of both raw data (at level) and
exchange/ tweets rate returns defined by $y'=100
\log{y(t)}-\log{y(t-1)}$ and $u'=100 \log{u(t)}-\log{u(t-1)}$ ,
respectively. Data differentiation accounts the problem of
non-stationarity and trends, thus eliminating potential biases in
forecasting.

We evaluated the forecasting performance of the above models on both
kind of test sets (at level and differentiated), through
(a)fixed-forecasting-horizon metrics, and, (b) trading simulations.
In particular, we used three fixed-forecasting-horizon metrics:
\\
(i)the root mean square error metric defined by\\
\begin{equation}
RMSE=\sqrt{\frac{1}{N_V}(\sum_{t=1}^{N_V} e(t)^2)}
\end{equation}
\\
(ii) the mean absolute error defined by
\begin{equation}
MAE=\frac{1}{N_V}\sum_{t=1}^{N_V} \vert e(t) \vert
\end{equation}\\

where $e(t)=\hat{y}(t)-y(t)$, $\hat{y}(t)$ is the prediction and
$y(t)$ is the actual closing rate at time $t$; other metrics such as
the mean square error were also used leading to the same
conclusions. \\
(iii) directional change statistics, namely

(a) for the analysis of the actual at level data, the average number
of ups and downs which are correctly
forecasted, defined by\\

\begin{equation}
DA=\frac{1}{N_{V}}\sum_{i=1}^{N_V} a(t)
\end{equation}
\\
where
\\
\begin{equation}\label{Multi}
a(t)= \left\{ \begin{array}{cc}
         1,\ if \ (y(t)-y(t-n_k))(\hat{y}(t)-y(t-n_k)) > 0\\
        0,\ otherwise \end{array} \right\}  \\
\end{equation}
\\

(b) for the analysis of the return rates (log-differentiated data),
the average number of signs that are correctly
forecasted defined by\\

\begin{equation}
Sgn=\frac{1}{N_{V}}\sum_{i=1}^{N_V} b(t)
\end{equation}
\\
where
\\
\begin{equation}\label{Multi}
b(t)= \left\{ \begin{array}{cc}
         1,\ if \ (\hat{y'}(t)y'(t)) > 0\\
        0,\ otherwise \end{array} \right\}  \\
\end{equation}
\\
Our trading simulations involved the computation of the return
profits defined by,

\begin{equation}\label{Multi}
R(t)=\sigma(t) \frac{y(t)-y(t-n_k)}{y(t-n_k)} 100\\
\end{equation}

Here we used the simple mving average trading rule reading:
\\
\begin{equation}\label{Multi}
\sigma(t)= \left\{ \begin{array}{cc}
         1,\ if \  \hat{y}(t) > \overline{y(t)}_{m}\\
        -1,\ if \  \hat{y}(t) < \overline{y(t)}_{m}\end{array} \right\}  \\
\end{equation}
\\
where

$\overline{y(t)}_{m}$ is the m-order moving average defined as
\\
\begin{equation}\label{Multi}
\overline{y(t)}_{m}=\frac{1}{m} \sum_{i=0}^{m-1} y(t-i)\\
\end{equation}

\section{Time Series Analysis and Trading Simulations: Results and Discussion}
\label{sec:3}Regarding the at level time series, the values of the
parameters of the AR models and their standard deviations as
obtained for different values of $n_a$ and $n_k=1$ are given in
table 1.

\begin{table}
\caption{At level data series: Coefficients of the AR models fitted
with the training dataset}
\label{tab:1}       
\begin{tabular}{lllllllllll}
\hline\noalign{\smallskip}
 & $a_1$ & $a_2$ & $a_3$ & $a_4$ & $a_5$ & $a_6$ & $a_7$ & $a_8$ & $a_{9}$ & $a_{10}$\\
\noalign{\smallskip}\noalign{\smallskip}
$n_a=1$ & -1 &  &  &  &  &  &  & &\\
$n_a=2$ & -1.0047 & 0.0048 &  &  &  &  &  & & \\
$n_a=3$ & -1.0014 & -0.0875 & 0.08914 &  &  &  &  & & \\
$n_a=4$ & -1 & -0.0802 & 0.1044 & -0.024 &  &  &  & & \\
$n_a=5$ & -1 & -0.0761 & 0.1013 & -0.0646 & 0.0397 &  &  & & \\
$n_a=6$ & -0.9976 & -0.078 & 0.1068 & -0.0687 &  -0.022 & 0.0598 &  & & \\
$n_a=7$ & -1.0014 & -0.0768 & 0.11010 & -0.0753 & -0.0172 & 0.1209 & -0.0601  & & \\
$n_a=8$ & -0.9934 & -0.0923 & 0.1119 & -0.0648 & -0.0339 & 0.1321 &0.0719  & -0.1312 & \\
$n_a=9$ & -0.9997 &  -0.0895 & 0.1199 & -0.0669 & -0.0314 & 0.1341 &0.0693  & -0.1852 & 0.0496\\
$n_a=10$ & -0.995 & -0.0997 &  0.1221 & -0.0597 &  -0.0335 & 0.13512 & 0.0723  & -0.1875 & 0.00297 &0.0429\\

\noalign{\smallskip}\hline
\end{tabular}
\end{table}

Figures 3a,b,c summarize the resulting $RMSEs$, $MAEs$, $DAs$ for
the AR models with respect to $n_a$ and $n_k=1$.\\
Based on the $RMSE$, $MAE$, we found that for one-step time
forecasting horizons the random walk model defined by $a_1=-1,
a_2=...a_{n_a}=0$ cannot be outperformed by any other AR model
(Figure 3a,b). The $RMSE$ for the random walk model was 0.00186, the
$MAE$ was 0.0013; the variance of the prediction error distribution
was $\approx 3.45E^{-6}$. Any other AR model with $n_a>1$ gave
greater or almost equal $RMSEs$, $MAEs$ than the ones obtained with
the random walk model. Incorporating now the information from the
Twitter's database in the ARX models (defined by $n_a>0$,$n_b>0$) we
constructed the 2-dimensional contour plots of the computed $RMSE$,
$MAE$ and $DA$ (Figures 3d,e,f) for $n_k=1$ and $n_a$,$n_b$ ranging
from 1 to 10. It is shown that the best ARX predictors were in the
range of $n_a=1,2$ and $n_b=2-10$ giving $RMSEs$ around 0.00181,
$MAE$ around 0.00128, $DA$ around 0.65; the variances of the
estimation errors were around $3.8E^{-6}$. The simulations results
indicate that no ARX model could significantly outperform the random
walk model in terms of the $RMSE$ and $MAE$. In fact, the apparent
best ARX predictor with $n_a=2$,$n_b=10$ gave a $RMSE$ equal to
0.00181 which is slightly better than the one obtained with the
naive random walk. However, the one-way analysis of variance
statistical test for the mean of the distribution of estimation
errors between the random walk and the best ARX predictor showed no
significant difference. In terms of the $DA$ metric though, it is
shown that the ARX models with $n_a=1,2$,$n_b=1-8$ resulted to
significant higher values compared to the one of the random walk
(see Figure 3f). In particular, for this range of parameters the
$DA$ ranged from 0.67 (for $n_a=1$,$n_b=1$) to 0.6 (for
$n_a=1$,$n_b=7$). The values of the ARX coefficients as well as
their uncertainty (standard deviation) for $n_a=1$ and $n_b=7$ are
given in Table 2.

\begin{table}
\caption{At level data: Coefficients and their standard deviation of
the ARX model with $n_a=1,n_b=7$}
\label{tab:2}       
\begin{tabular}{llll}
\hline\noalign{\smallskip}
\noalign{\smallskip}\hline\noalign{\smallskip}
$a_0=0.97357(\pm0.0099)$ & $b_0=0.0334(\pm0.0132)$ & $b_1=-0.0265(\pm0.0153)$\\
$b_2=0.01634(\pm0.0152)$ & $b_3=-0.00388(\pm0.01498)$ & $b_4=-0.00309(\pm0.01495)$\\
$b_5=0.00622(\pm0.0149)$ & $b_5=-0.0172(\pm0.0149)$ & $b_6=0.02099(\pm0.0128)$\\

\noalign{\smallskip}\hline
\end{tabular}
\end{table}

Similar results with the above were obtained using the ANN models.
Figures 3g,h,i summarize the corresponding $RMSEs$, $MAEs$ and
$DAs$. The best ANN predictors were found for $n_a=1$ and $n_b=1,2$
with $RMSE$ around 0.0017, $MAE$ around 0.00123 and $DA$ around
0.65. For this range of parameters the variances of the estimation
errors were around $3E^{-6}$).\\
The above results indicate that the information contained in the
Twitter could be used to enhance the forecasting efficiency in the
short (intradaily) run.

\begin{figure*}
  \includegraphics[width=1.25\textwidth]{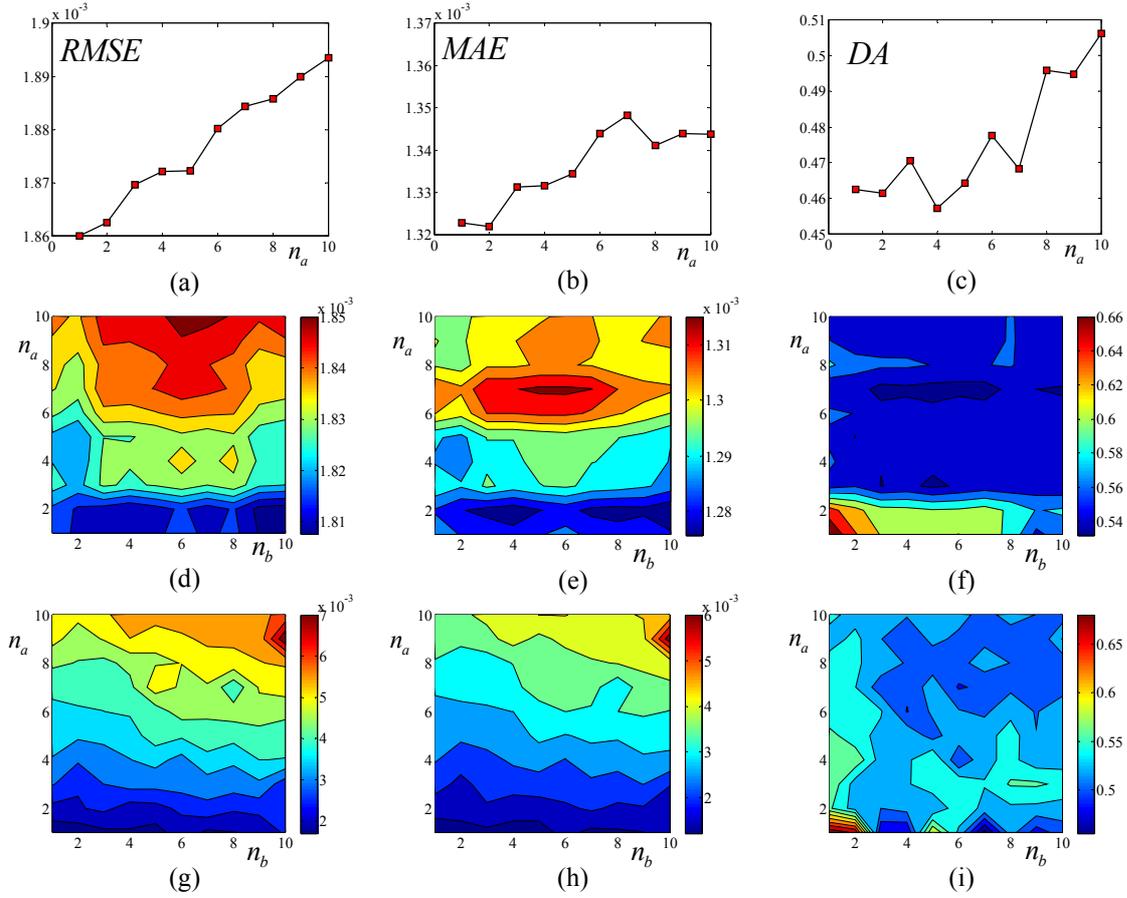}
\caption{At level data analysis: $RMSEs$((a),(d),(g)),
$MAEs$((b),(e),(h)) and $DAs$ ((c),(f),(i)) for the AR, ARX and ANN
models respectively with respect to the orders $n_a$ and $n_b$ with
$n_k=1$}
\label{fig:3}       
\end{figure*}
We also performed computations with other forecasting horizons
defined by $n_k>1$. For illustration purposes, Figures 4,5 summarize
the $RMSEs$, $MAEs$ and $DAs$ for $n_k=2$ and $n_k=4$, respectively
as computed with AR, ARX and ANN models. The corresponding variances
are around $1E^{-5}$ for both ARX and ANNs models. As it is shown
the ARX and ANN models outerperform the naive random walk with
respect of all metrics when $n_k>1$. However, this should be
attributed to the apparent trend in the actual/raw at level data. It
is interesting though to remark, that even at relatively long
forecasting time horizons (e.g. for $n_k=4$) the information
contained in the twitters enhances significantly the forecasting
performance(see e.g. Figures 5f,g,h).

\begin{figure*}
  \includegraphics[width=1.25\textwidth]{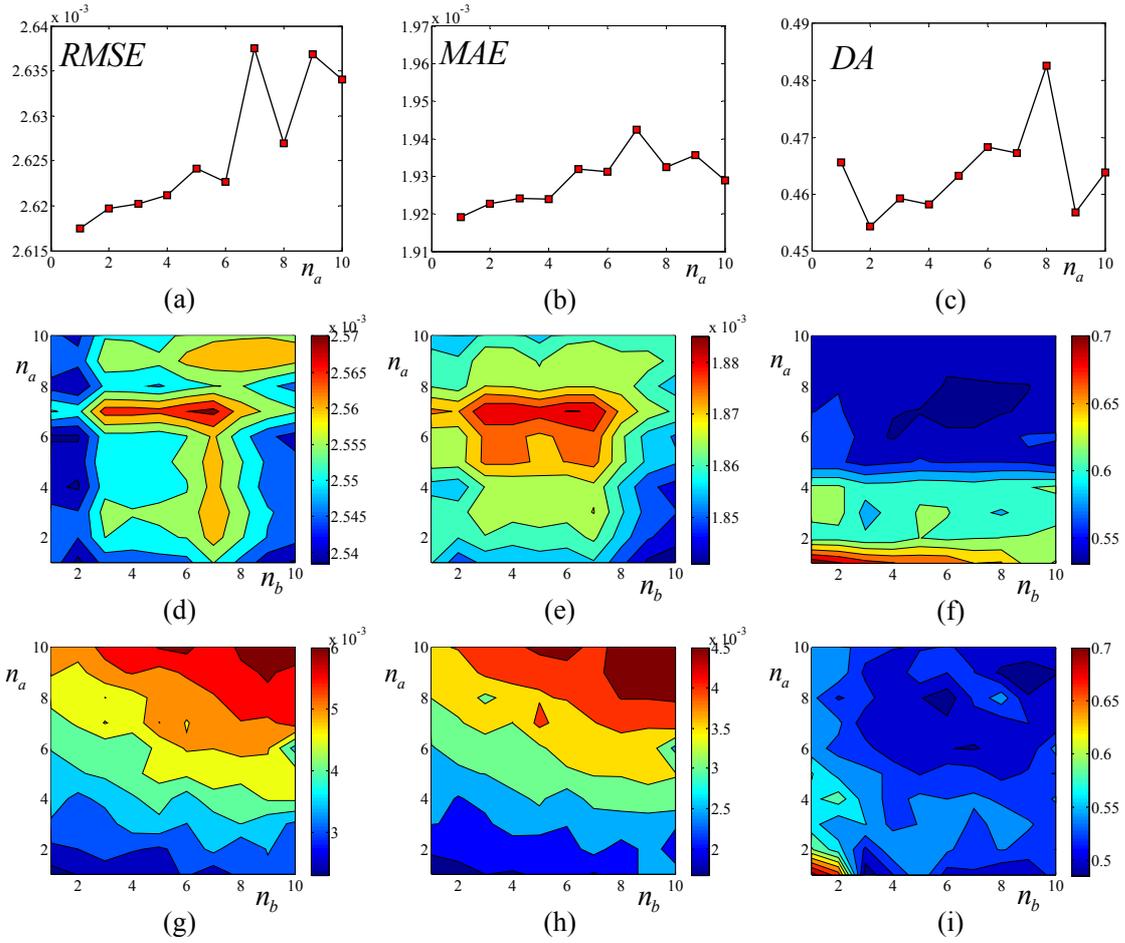}
\caption{At level data analysis:  $RMSEs$((a),(d),(g)),
$MAEs$((b),(e),(h)) and $DAs$ ((c),(f),(i)) for the AR, ARX and ANN
models respectively with respect to the order $n_a$ and $n_b$ with
$n_k=2$}
\label{fig:4}       
\end{figure*}

\begin{figure*}
  \includegraphics[width=1.25\textwidth]{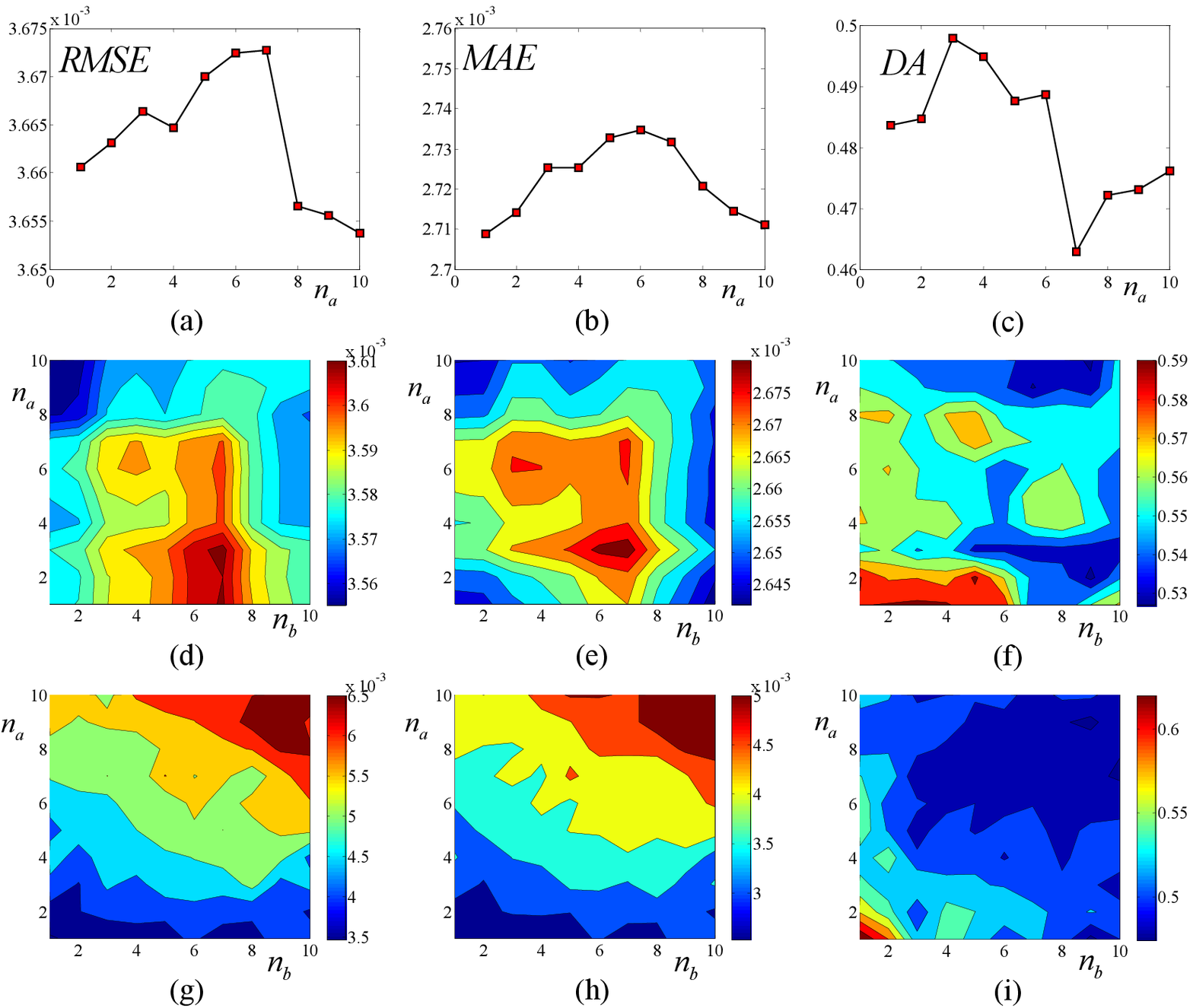}
\caption{At level data analysis:  $RMSEs$((a),(d),(g)),
$MAEs$((b),(e),(h)) and $DAs$ ((c),(f),(i)) for the AR, ARX and ANN
models respectively with respect to the orders $n_a$ and $n_b$ with
$n_k=4$}
\label{fig:5}       
\end{figure*}

We also performed trading simulations in which the ``traders" use
the estimated price  as obtained by the forecast of the models and
produce a ``buy" signal ($s = 1$) if the estimation is above the
current moving average actual closing, and a ``sell" signal $(s =
-1)$ otherwise as described in the previous section. Figure 6a shows
the cumulative return of the random walk model for $m=4$, $nk=1$. As
it is shown the trading simulations result to cumulative loses.

\begin{figure*}
  \includegraphics[width=1.0\textwidth]{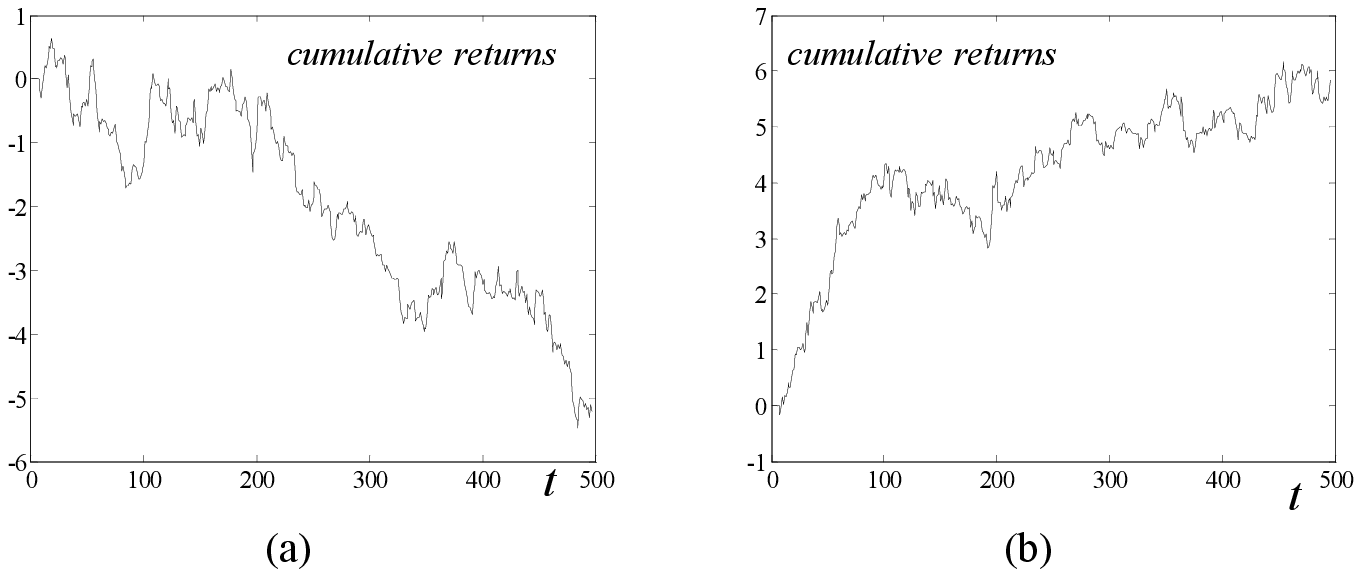}
\caption{Cumulative returns derived with the moving average trading
simulation (with moving average horizon defined by $m = 4$) for a
one step forecasting horizon ($n_k=1$) using (a) the random walk
model, (b) a ANN model with $n_a=3$ and $n_b=6$}
\label{fig:6}       
\end{figure*}

Figure 7a illustrates the mean values of the computed cumulative
returns for $m=4$, $nk=1$ when using the AR models. Figures 7b,c
show the contour plots of the mean values of the cumulative returns
computed with the trading simulations for $m=4$, $nk=1$ with the ARX
and ANN models, respectively. The trading simulations, indicate that
the tweets incorporated in the ARX and ANN models carry information
that enhances the forecasting ability resulting, for certain values
of the orders $n_b$ and $n_a$, into profitable trading opportunity,
thus outerperforming the AR models (which lack information from
tweets). Indicatevely, in Figure 6b we illustrate the cumulative
returns obtained by the moving average trading simulation for
one-step forecasting horizon ($n_k=1$) using a ANN model with
$n_a=3$ and $n_b=6$.

\begin{figure*}
  \includegraphics[width=0.75\textwidth]{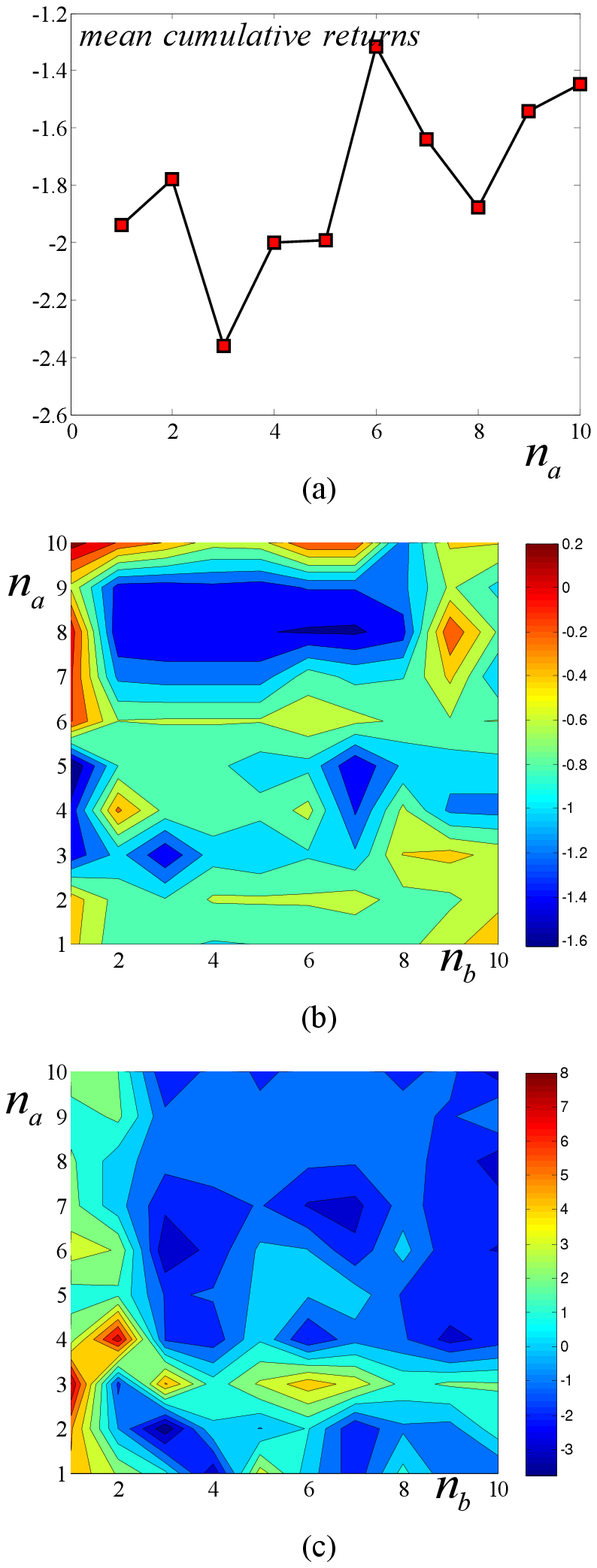}
\caption{Mean Cumulative returns derived with the moving average
Trading simulation ($m = 4$ and $n_k = 1$) using the (a) AR, (b)
ARX, (c) ANN models}
\label{fig:7}       
\end{figure*}

 At this point we should note that the above results are
due to trending in the at level data. To test if the forecasting
efficiency of the ARX and ANN models employing information contained
in the twitter's database perform  better than the naive random walk
and generally the AR models when trend is excluded, we also
performed the same analysis using the detrended data as derived by
log-differencing. Figures 8,9,10 depict the computed $RMSEs$,$MAEs$
and $Sgns$ for the AR, ARX and ANN models for the exchange/ tweets
rate returns (log-differences), for $n_k=1$, $n_k=2$ and $n_k=4$,
respectively (the orders $n_a$,$n_b$ ranged from 1 to 10). For
$n_k=1$ the random walk gives the minimum $RMSE$ and $MAE$ (around
0.1405 for $RMSE$ and 0.1 for $MAE$) compared to the other AR
models. For the same time lag, the best ARX model
($n_a=1$,$n_b=8-10$)gave around 0.1372 for $RMSE$ and 0.0975 for
$MAE$. The corresponding $RMSE$ and $MAE$ values of the best ANN
model ($n_a=1$,$n_b=2$ were 0.1361 for $RMSE$ and 0.098 for $MAE$.
One-way Anova test between the mean values of the estimation errors
from the best ARX and ANN models and the random walk showed no
significant difference. The same behaviour is observed for other
$n_k>1$. For example, for $n_k=4$ the random walk gives the minimum
$RMSE$ and $MAE$ (around 0.140 for $RMSE$ and 0.1 for $MAE$)
compared to the other AR models. For the same time lag, the best ARX
model ($n_a=1$,$n_b=6$)gave around 0.1395 for $RMSE$ and 0.1 for
$MAE$. The corresponding values of the best ANN model
($n_a=6$,$n_b=6$ were 0.1361 for $RMSE$ and 0.0986 for $MAE$. Again
one-way Anova test between the mean values of the estimation errors
from the best ARX and ANN models and the random walk showed no
significant difference.In terms of directional change statistics s
described by the $Sgn$  describing the proportion of times that the
model forecasts correctly the sign of change in the rates, the AR
models for $n_k=1$ result to values of $Sgn$ around 0.5 (the random
walk gives around 0.53). However, it is interesting to note that
some ARX and ANN models incorporating the information contained in
the tweets produced considerably higher values of $Sgn$ up to 0.59
for (see Figure 8f,i).

\begin{figure*}
  \includegraphics[width=1.25\textwidth]{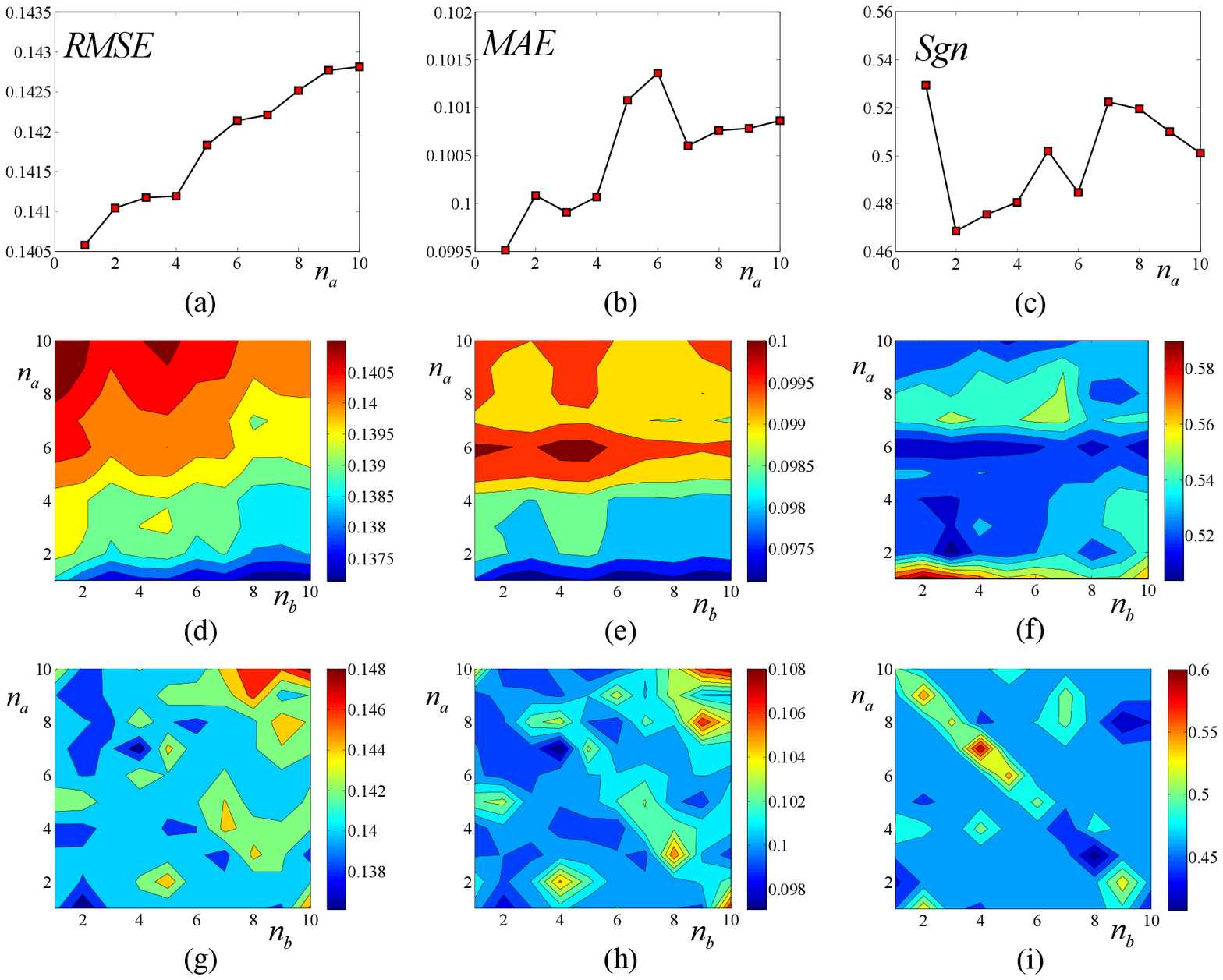}
\caption{Rate returns (log-differences)analysis:
$RMSEs$((a),(d),(g)), $MAEs$((b),(e),(h)) and $Sgns$ ((c),(f),(i))
for the AR, ARX and ANN models respectively with respect to the
orders $n_a$ and $n_b$ with $n_k=1$}
\label{fig:8}       
\end{figure*}

\begin{figure*}
  \includegraphics[width=1.25\textwidth]{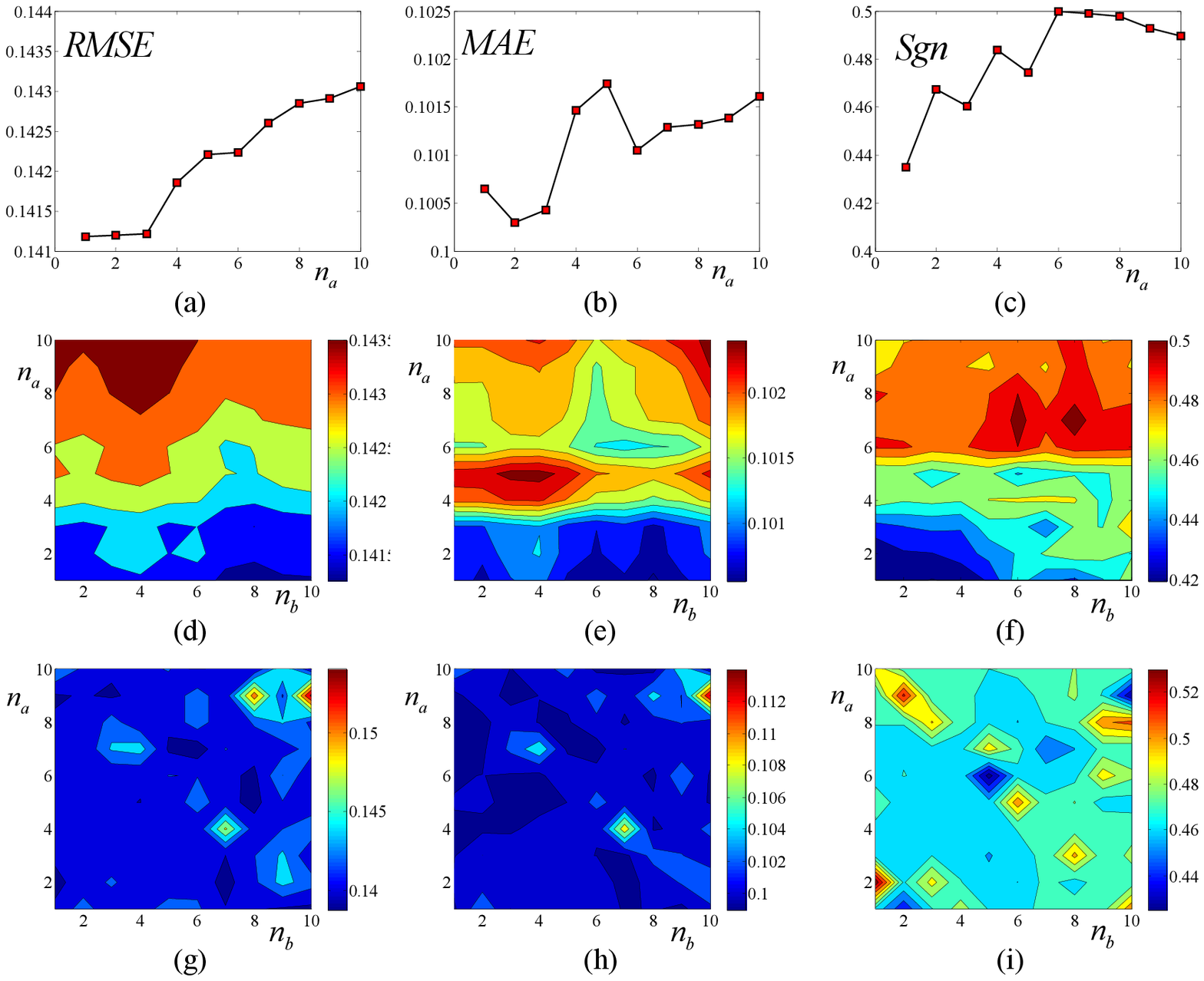}
\caption{Rate returns (log-differences) analysis:
$RMSEs$((a),(d),(g)), $MAEs$((b),(e),(h)) and $Sgns$ ((c),(f),(i))
for the AR, ARX and ANN models respectively with respect to the
orders $n_a$ and $n_b$ with $n_k=2$}
\label{fig:9}       
\end{figure*}

\begin{figure*}
  \includegraphics[width=1.25\textwidth]{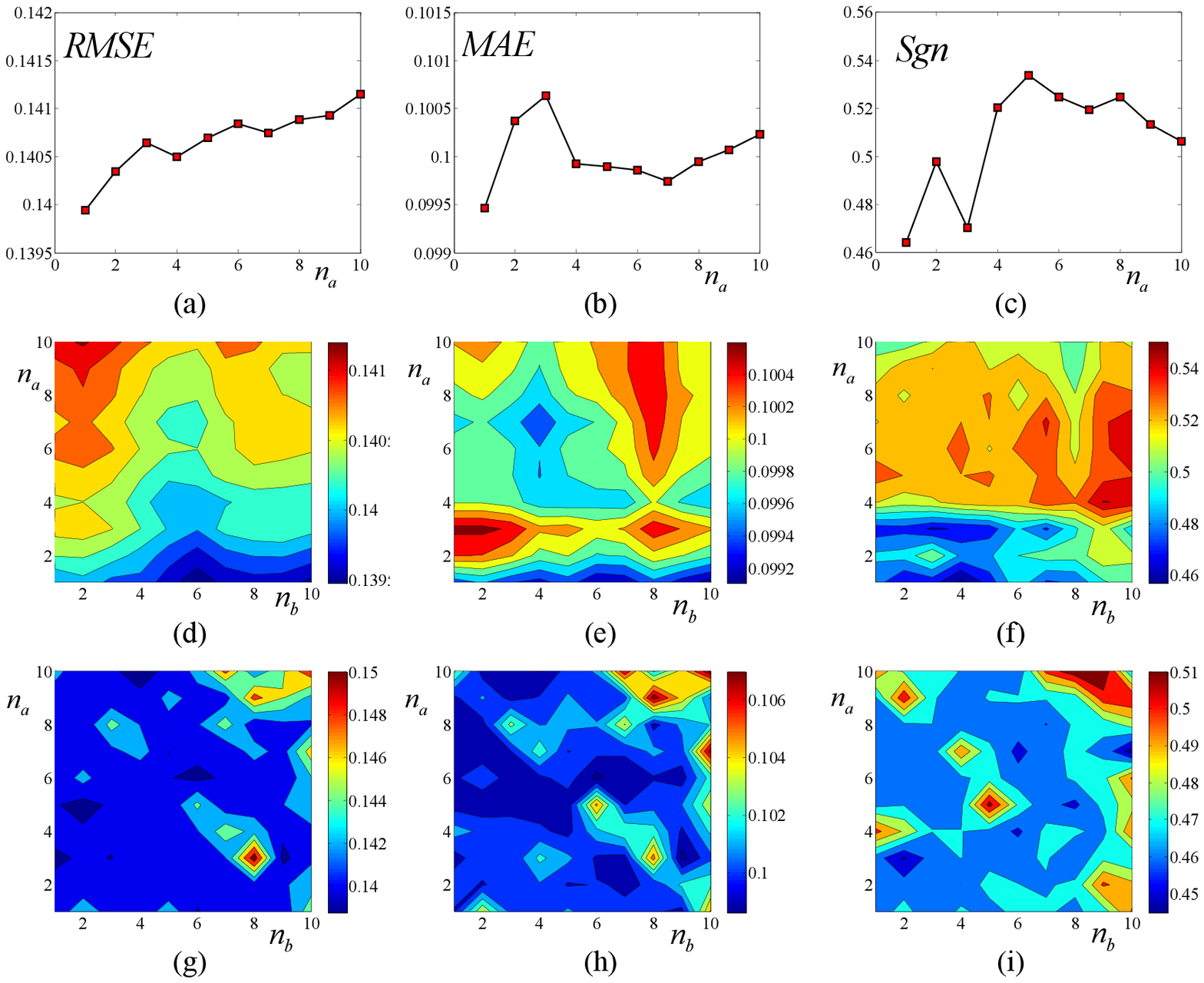}
\caption{Rate returns (log-differences) analysis:
$RMSEs$((a),(d),(g)), $MAEs$((b),(e),(h)) and $Sgns$
((c),(f),(i))for the AR, ARX and ANN models respectively with
respect to the orders $n_a$ and $n_b$ with $n_k=4$}
\label{fig:10}      
\end{figure*}

In particular, the best ARX models, with respect to the $Sgn$
statistic, was found for $n_a=1$, $n_b=1-4$, $n_k=1$ giving values
of $Sgn$ from 0.574 ($n_a=1$, $n_b=4$) up to 0.596 ($n_a=1$,
$n_b=2$) (Figure 8f). The best ANN performance is observed for
$n_a=7$, $n_b=4$ and $n_k=1$ giving a value of $Sgn$ around 0.6 (see
Figure 8i). In order to test the statistical significance of the
results produced by the best ARX and ANN models we performed
bootstrapping on a total of 5000 randomly perturbed resamples of the
validation data. Simulations were performed for $n_k=1$ using the
best ARX model (with $n_a=1$, $n_b=2$) and the best ANN model (with
$n_a=7$, $n_b=4$). The resulting empirical bootstrap distributions
of the $Sgns$ obtained by the ARX and ANN models on the 5000
resamples are illustrated in Figure 11a,b, respectively.

\begin{figure*}
  \includegraphics[width=1.25\textwidth]{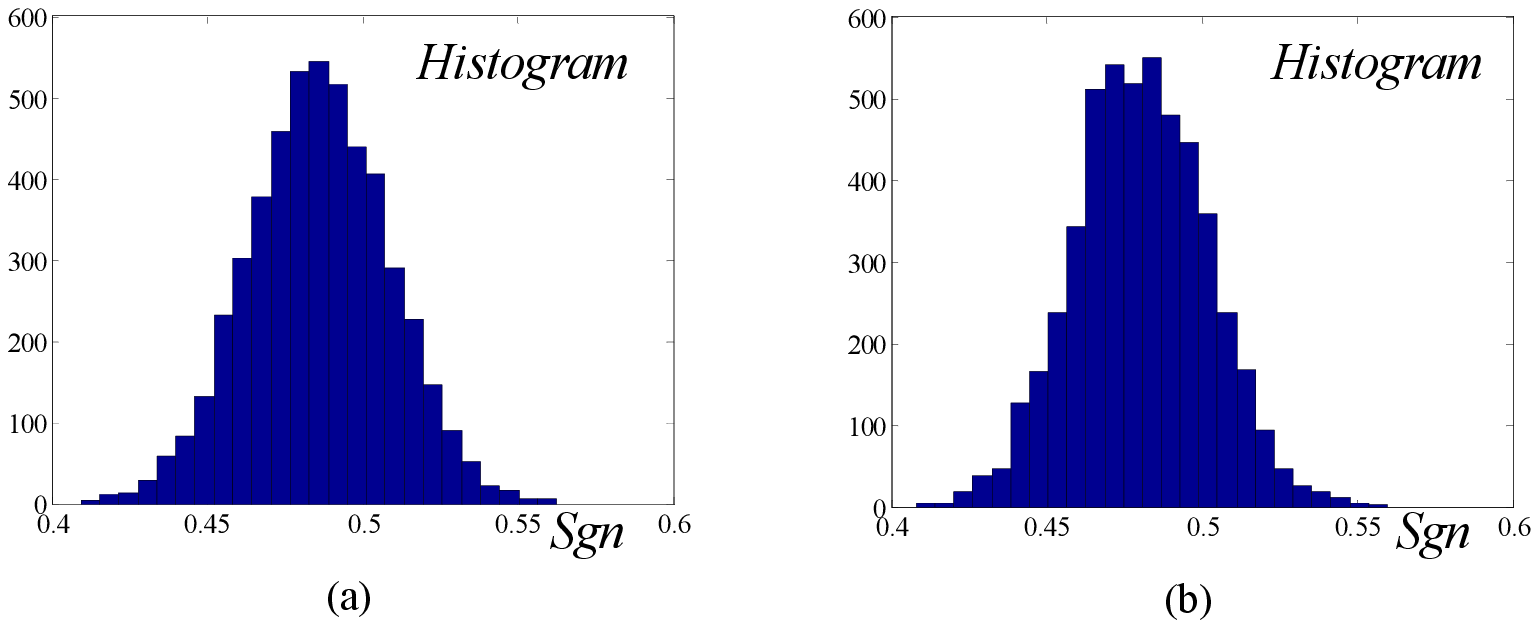}
\caption{Empirical Bootstrapping distributions of $Sgns$ for $n_k=1$
using (a) the ARX(with $n_a=1$, $n_b=2$), (b) the ANN (with $n_a=7$,
$n_b=4$ produced from 5000 bootstrapped resamples of the validation
data}
\label{fig:11}       
\end{figure*}

As it is clearly seen the values obtained with the best ARX and ANN
models are well beyond the maximum values of the resulting bootstrap
distributions. These results indicate that the information of the
tweets can be used to enhance the forecasting efficiency of the
directional change of the rates. For large values of the forecasting
horizon i.e. for $n_k>1$ no significant differences were observed
(the values of $Sgns$ were around 0.5 with small deviations) (see
Figures 8c,f,i and Figures 9c,f,i)

\section{Conclusions}
\label{sec:4} Over the last years it has been demonstrated that
social media such as web-based search engines and recently Twitter
can be used to forecast certain future complex events. In a similar
fashion, an unresolved problem in contemporary monetary and
economical management research, the foreign currency exchange rate
forecasting (forex) in the short run, might be also benefit from the
use of social microblogging. In this work we attempted to test the
validity of the Efficient Market Hypothesis in its strong form with
respect to forex through Twitters' social networking communication
platform. We explored the possibility of using private (yet,
publicly available through a microblogging platform) information of
market players that could be used to outerperform the EMH in the
very short term dictating that the market is inefficient, as far as
the overall information flow to investors is concerned. In
particular, we attempted to give an answer to the following basic
question: can information behavior contained in the context of
microblogging and the ``belief" of traders be used to enhance the
forecasting efficient and outperform the random walk? To our
knowledge this is the first time that such an analysis is provided
for the forecast of the exchange rate of EUR/USD. We believe that
our study is of significant importance, as the contemporary research
in the area trying to contradict the EMH is focused on the
uncovering of market anomalies as these may arise by the information
flow provided by the market players' ``beliefs" in the social
networks. Towards this direction, the development of several
psychological indexes that are related to market's consensus, is
already in great use by players as well as market's regulators. Our
analysis showed that the rate exchange forecasting at level, based
on people's beliefs as these can be data-mined through microblogging
may carry significant information that can used to outperform the
random walk hypothesis and AR models that do not include such
information but solely past values of the exchange rates, in the
very short run. This was also demonstrated through moving average
trading simulations. However, we should note that this behaviour
should be attributed to the underlying trend of the data. This is
apparent when encountering larger forecasting horizons. When the
analysis was performed on the return rates (log-differencing), i.e.
a log-differencing of the actual values which accounts the problem
of non-stationarity and trends, the analysis showed significant
difference with respect to forecasting efficiency of directional
changes as described by the proportion of times that the relative
directional changes of signs are correctly forecasted for
one-step-time forecasting horizon. Regarding any conclusions that
can be extracted by our analysis, we should refer to its certain
assumptions and restrictions. For example we used (a) a data-base
deploying within a limited period of time that did not include any
major anomaly, (b) our forecasting was targeted solely in the very
short (intradaily) horizon, (c) no risk assessment analysis was
taken into account, (d) we used just black-box time series models,
(e) we used only a small part of the social microblogging platforms
reported data. In addition, evaluating forecasting efficiency and
accuracy remain an important issue for further research. For
example, out-of-sample statistical measures can be improved using
rolling-origin evaluations and re-calibration of optimal
coefficients based on new data sets (see Tashman (2000) for a review
and critical discussion). Concluding, we believe that social
networks can provide the basis for further advances in the field and
thus enable the formalization of the experimental side of the
market's psychology as this is shaped by the human behavior. Towards
this aim detailed Agent-based models analysed by state-of-the-art
multiscale techniques (see e.g. Tsoumanis et al. (2010), Siettos et
al.(2012))have the potential to facilitate computational modeling
and exploration - and thus our understanding and forecasting
market's complex dynamics.




\end{document}